# New Method for Injectable Quinine Quality Assurance Control Using a Multi-Spectral Microscope

**A. Sow[1], I. Traore[1], T. Diallo[2,3], A. BA[1]**

[1]Laboratoire d'Optique, de Spectroscopie et des Sciences Atmosphériques (LOSSA), Faculté des Sciences et Techniques (FST), Université des Sciences, des Techniques et des Technologies de Bamako, Bamako, Mali
[2]Département des Sciences du Médicament, Faculté de Pharmacie, Université des Sciences, des Techniques et des Technologies de Bamako, Bamako, Mali
[3]Laboratoire National de la Santé (LNS), Bamako, Mali
Email: amenotra@lossa-mali-edu.org

## Abstract

Africa is the world region that is most affected by malaria. Among the therapies used, injectable quinine is considered to be one of the effective antimalarial drug, however non-quality assured antimalarials clearly have a strong market penetration across Africa. To overcome this problem, it becomes more and more necessary to set up quantitative and qualitative analysis system for antimalarial quality control. The objective of the present investigation is an attempt to use customized multispectral microscope equipped with UV-Visible lasers for injectable quinine quality assurance control routinely. For that, we have established the calibration curve of quinine solution concentration as a function of area under light intensity histogram crossing the solution. From this calibration curve, we can check the conformity of any injectable quinine according to the pharmacopoeia involved. The proposed technique is a promising alternative for drug quality control routinely.

## 1. Introduction

Malaria remains one of the most important infectious diseases in the world with an estimated 212 million cases and 429,000 deaths resulting occurred worldwide. Most of the both infectious and deaths cases in 2015 were in the WHO African Region (90% and 92%) [1]. A study carried out in Ivory Coast on Occupational nand economic impacts of falciparum malaria in a private sector company showed that Morbidity and socio-economic impact of malaria in the workplace remain high despite the Abuja call [2].

In Mali, the 2015 health statistics show 2.369245 cases of malaria, including 686,017 severe malaria and 1978 deaths, that representing a case-fatality rate of 8.9%. At the economic level, malaria affects Mali's annual economic growth by 1.3% due to absenteeism at work or school. According to a study conducted by the National Institute of Public Health Research (INRSP), economic losses amount to 72 billion CFA francs [3]. Thereby, malaria has become a significant obstacle to the economy of endemic country which is in general developing countries. Fortunately, several solutions could be used to fight malaria such as prevention methods, treatments and vector control. In this study we will focus in antimalarial drugs, particularly injectable quinine despite the fact of injectable artesunate is recommended in many studies [4] [5] [6] [7] for the treatment of severe malaria. Moreover, in most Africa countries quinine is still used as monotherapy and remains the first line drug contrary to the WHO recommendations [8] [9]; this is why quinine continues to play a significant role in the management of malaria in sub-Saharan Africa and other malaria endemic areas, and its use in routine practice may not be restricted to the stated WHO recommendations according to Achan *et al.* [10].

Several studies [10] [11] [12] [13] have highlighted the failure of quinine treatment resulting in the large part from quinine resistance, incorrect dosing, irrational prescription practice, poor patient compliance and mainly the high prevalence of counterfeit drugs which affect considerably the quality of quinine used in routine care.

Indeed, a study conducted by Vicki Brower *et al.* [14] examining the quality of malaria drugs in eight countries earlier this year showed an alarmingly large market of substandard and counterfeit drugs in sub-Saharan Africa, in both private and public. Non-quality assured antimalarial clearly have a strong market penetration across Africa. The same study states that counterfeit or substandard drugs account for 10% of the world's pharmaceutical market; however, in Africa the proportion is about 30%, and the consequences can be deadly for malaria, especially for pregnant women and children. In spite of enhanced control efforts, counterfeit medicine is become a major public health problem. It is therefore necessary that the different agencies in charged for drug control and law enforcement in countries and at the international level, act in close collaboration.

In addition to repressive measures, human resources and quality control techniques for drugs should be strengthened. For good treatment, medications must be safe, effective and of consistent quality to produce the expected therapeutic effect. Thus, according to World Health Organization, quality control refers to the sum of all procedures undertaken to ensure the identity and purity of a particular pharmaceutical. Such procedures may range from the performance of simple chemical experiments which determine the identity and screening for the presence of particular pharmaceutical substance, to more complicated require-ments of pharmacopoeia monographs [15].

In Mali apart from the National Laboratory of the Health (LNS), there is no adequate structure which can determine the quality assurance of antimalarial drug (drug in general).

To overcome this situation and contribute to combat counterfeit drugs, we have conducted the present study whose general objective is to study the possibility to establish a quality control system routinely.

Many technologies could be used to detect counterfeit antimalarial drug such as Raman spectroscopy, liquid chromatography, mass spectrometry, Fourier-transform infrared spectroscopic imaging, colorimetric assays, and Near-infrared spectroscopy [16]. The main goal of this study is to highlight the possibility to use a customized microscope to test specifications for the amount of active drug ingredients according to pharmacopoeia involved. The microscope used is a horizontal microscope designed by African Spectral Imaging Network (AFSIN) and made available to laboratories members of this network.

This microscope has the particularity to captures images and spectra in transmittance and reflectance modes.

## 2. Experimental Setup and Methods

### 2.1. Sample Preparation

#### 2.1.1. Solvent Solution Preparation

We have prepared a 1 N concentration of HCl by diluting 82.89 mL of hydrochloric acid in 1000 mL of distilled water. From this HCl solution, 10 mL was again dissolved in 1000 mL distilled water to prepare HCl solvent with concentration of 0.01 N. This last final solution of 0.01 N HCl is used as the solvent for quinine. The HCl solution is supplied by Batch: 10I140510 VWR (PROLABO$^*$).

#### 2.1.2. Preparation of Standard Stock Solution of Quinine

The quinine standard we used was provided by the Malian National Laboratory of the health.

Firstly, we have accurately weighed 10 mg of the standard solution of dihydrochloride quinine which was dissolved in a 100 mL volumetric flask with 0.01 N HCl acid.

Secondly, an aliquot of this stock standard solution was further diluted with adequate volumes of 0.01 N HCl to produce concentrations ranging from 0,075 to 0.3 $\mu gL^{-1}$. All the solutions prepared were homogenized in an ultrasonic apparatus for 2 to 5 minutes and then ready to be analysed by a multispectral microscope.

### 2.2. Imaging System

Images were acquired using a novel multispectral imaging system developed by Laboratoired' Instrumentation Image et Spectroscopie, INP-HB, DFR-GEE, Yamoussoukro, Côte d'Ivoire and presented in Ref. [17].

We have modified the original system by adding a sample holder system which consist a quartz cuvette with 1 cm path length (analysed solution container) and an input/output fiber optic cable to send transmission light to the objective of microscope. The sample is illuminated by monochromatic light source emitting in UV region (405 nm) in transmission, however, each wavelength available (RGB) can be used in transmission and also in reflection at any moment without modifying the system.

The lasers are controlled by a data acquisition card (DAQ) from National Instruments (USB 6008).

The transmission light issue from sample is viewed by a plan objective 15x/0.28 (a adjustable ReflX™ objective marketed by Edmund optics), then imaged with digital Lumenera lt225M camera. This camera has a pixel size of 5.5 × 5.5 μm and high resolution 2/3 CMOSIS CMV2000 sensor with a fully electronic global shutter. The laser, the camera and the motors that monitor the moving of the sample in X, Y, Z directions (note that in this experiment, sample is fixed as described in Figure 1) were controlled from a PC using a custom-made program in MATLAB software [18].

## 2.3. Image Acquisition and Analysis

The sample to be analysed (quinine in acid medium of concentration C) was placed in a quartz tube fixed in sample holder. The fraction of the incident UV light (405 nm) transmitted through the solution, then the objective of microscope produces an image that can be seen in Figure 2.

On the main screen, we can observe inside the circle a weak spot light and on the right-hand side on secondary screen we observe too the histogram representing the number of pixels as a function of light intensity. To access to the data of histogram, we have captured the image of main screen as a tiff file.

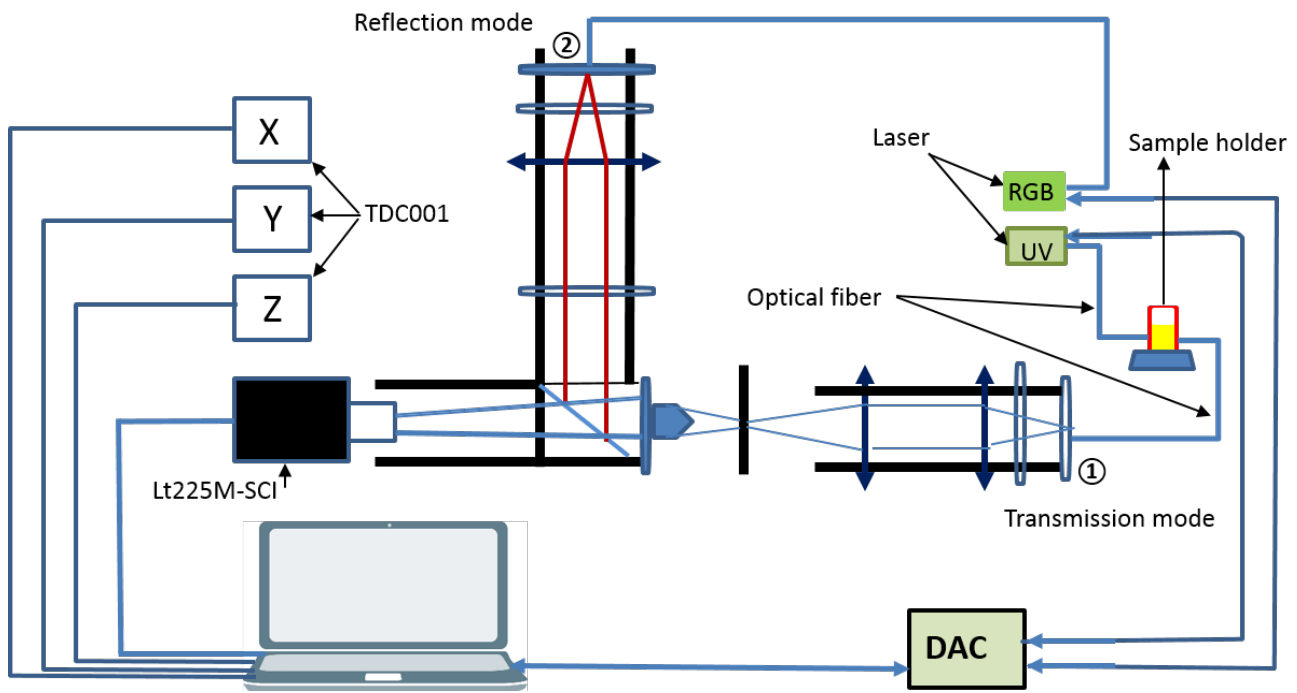

**Figure 1.** scheme of the multispectral microscope.

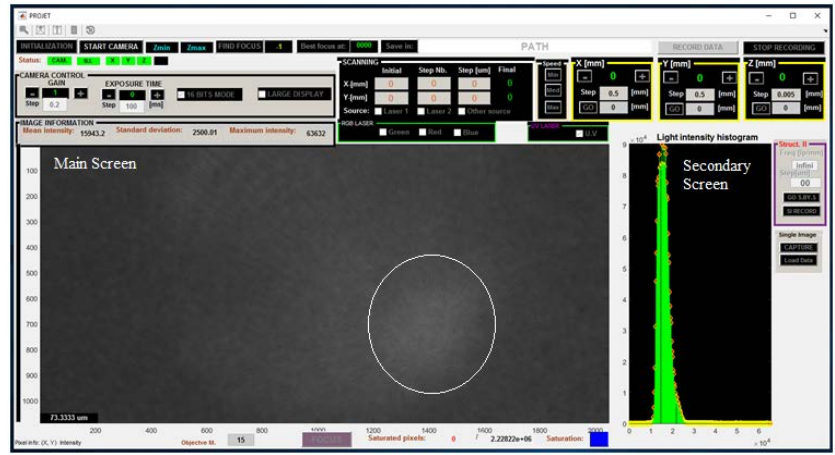

**Figure 2.** Screenshot showing light spot and histogram.

This tiff file is processed under MATLAB software to reproduce the histogram in the secondary screen of **Figure 2**. We then smoothed and adjusted all the histograms corresponding to the different quinine solution concentrations by appropriate functions and Gaussian with a correlation coefficient closest to 1in order to obtain their equations. According to these equations, we have determined the area under these histograms. Note that for each solution, the same acquisition parameters are used.

For instance, the curve of **Figure 3** was smoothed and then adjusted with a Gaussian function of order 4. The equation of this Gaussian is:

$Y_1 = 1.286e + 005 \times \exp\left(-((x - 107.3)/7.579)^2\right) - 5.028e + 004 \times \exp\left(-((x - 108.2)/5.361)^2\right) + 3.56e + 004 \times \exp\left(-((x - 98.81)/4.922)^2\right) + 3.442e + 004 \times \exp\left(-((x - 114.6)/10.92)^2\right)$ with a correlation coefficient $R^2 = 0.9999$.

Applying area under curve technique [19] [20], we have found: $A_1 = 2.2266e + 006$ a.u.

## 3. Results and Discussion

The area under the curve values for each quinine standard solution were determined as a function of the concentrations and the calibration graph was constructed as can be seen in **Figure 4**.

Considering that $A_0$ is the value of area under the curve of HCl solvent (c = 0.01 N), the ratio $\dfrac{A}{A_0} = T$ which expressed the transmittance is used to calculate absorbance with Beer-Lambert law $A_b = -\log T = -\log\left(\dfrac{A}{A_0}\right)$. Therefore, we obtain the calibration graph of absorbance as a function of concentration (**Figure 5**). In Mali quinine hydrochloride is available in 100 mg·ml⁻¹ preparation for injection. According to the Clarke's analysis of drugs and poisons [21], the percentage content of quinine should be varying between 95.0% and 105.0% of the nominal value X (X = value given on injectable vials), namely: 0.95X ≤ X ≤1.05X (1). To check the conformity of any sample of injectable quinine, 1 mL of this sample is taken and then dissolved in 100 ml of 0.01 N HCl. From this last solution, 1 ml is still pipetting then dissolved in 100 ml of 0.01 N HCl.

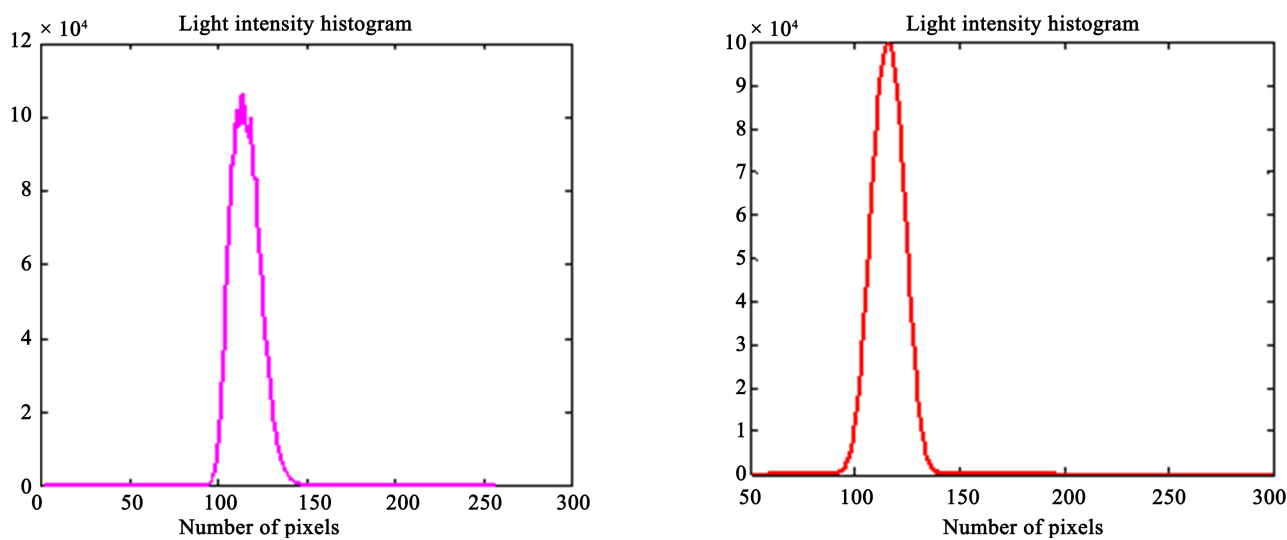

**Figure 3.** Luminous intensity curve of quinine standard solution of concentration c= 0.075 μgL⁻¹ adjusted (right) and unadjusted (left).

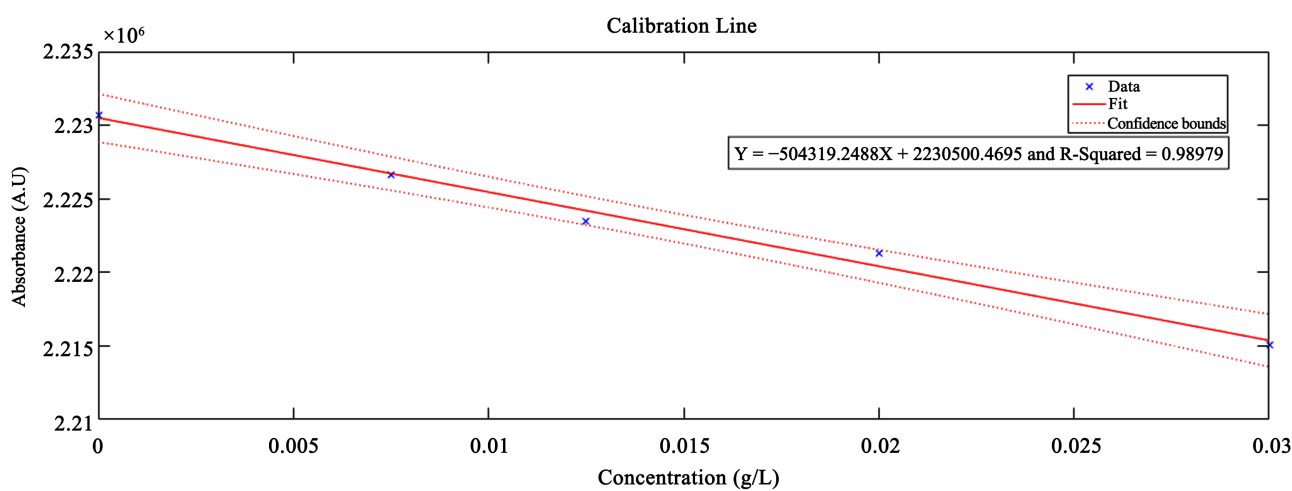

**Figure 4.** Graph of concentration versus area under intensity curve.

The area of the histogram corresponding to this solution is determined and the value injected into the equation $y = -504319.2488x + 2230500.4695$. The value of *x* obtained will be compared to that on the sample label with uncertainty within the limits defined in the standard British Pharmacopoeia and Clarke's (1). This method was validated by a test carried out on a sample of quinine that conformity was already established by National Laboratory of health of Mali with the UV spectrophotometric method.

In accordance with the data available on optical absorption spectrum of Quinine we have plotted molar extinction coefficient of quinine sulphate in 0.05 and 0.5 M $H_2SO_4$ [22] depending on wavelength (**Figure 6**). We notice that from 405 nm (the used wavelength), the molar extinction molar starts to diverge.

The value of 32 cm$^{-1}$/M obtained in the frame of this study with the standard solution of dihydrochloride quinine is under the values of 62 and 57 cm$^{-1}$/M of quinine sulphate in 0.05 and 0.5 M $H_2SO_4$ respectively for 405 nm. However, our value is in the same order of magnitude. This result strengthened our approach.

### The Limit of This Study

We observe in **Figure 5** a very low value of the absorbance. This observation is disputable since, according to Clarke's standard document, the peaks of the UV-visible absorption spectrum of quinine correspond to the wavelengths 250 nm, 326 and 346 nm. Despite the fact that our microscope does not have any of these wavelengths, this limitation cannot affect significantly the quality of our measurements.

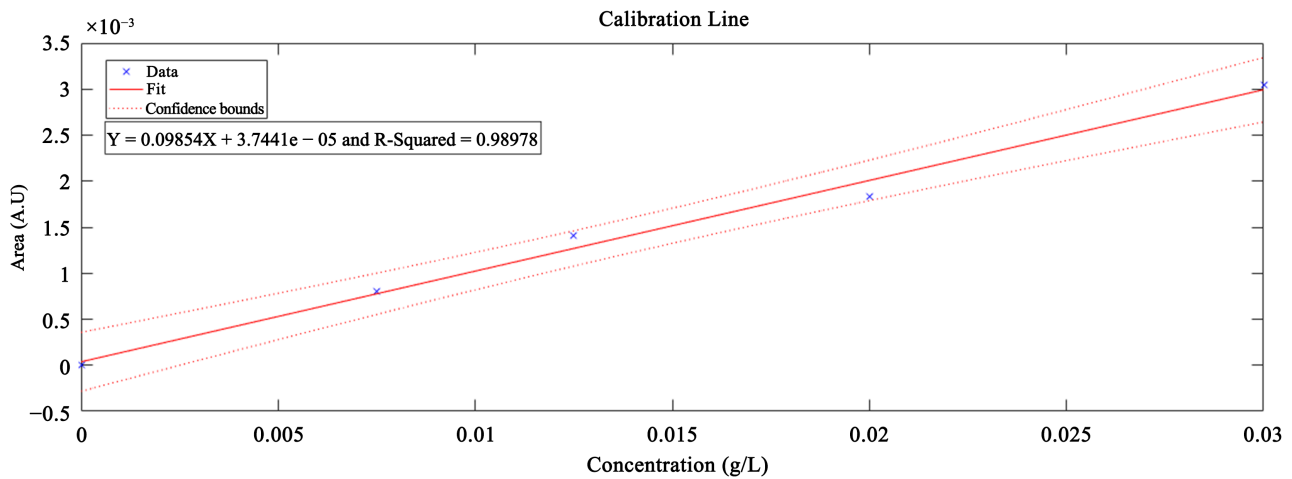

**Figure 5.** Calibration graph of absorbance from multiple quinine standard solutions of different concentrations.

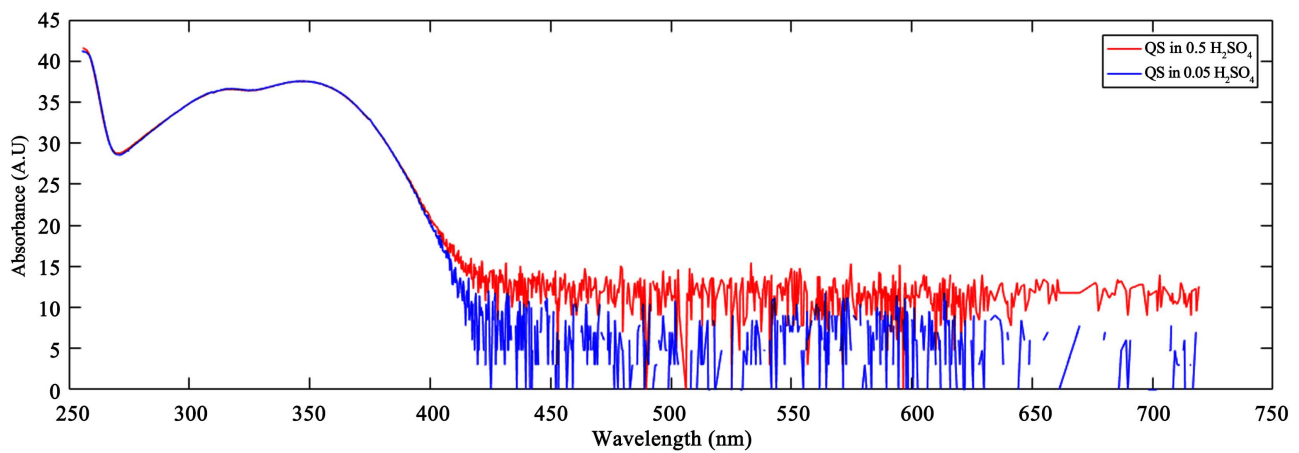

**Figure 6.** Molar extinction coefficient vs wavelength for Quinine Sulphate (QS) dissolved in 0.5 $MH_2SO_4$ (red) and 0.05 $H_2SO_4$ (blue).

## 4. Conclusion

Through this preliminary and prospective study, we have achieved our goal of showing that it is possible to use a multispectral microscope for injectable quinine quality assurance control in particular and drugs in general. The developed method comes to strengthen the UV spectrophotometric method already used by National Laboratory of the health of Mali for drug routinely control and enriched the existing techniques. The results obtained could be improved with adequate laser light sources. Moreover, further study could be based on the present approach.

## Acknowledgements

We would like to thank the International Science Program (ISP) for the funding and donation of microscopes through the Laboratory of Optics, Spectroscopy and Atmospheric Sciences (LOSSA) at the Faculty of Sciences and Techniques of Bamako. We gratefully acknowledge the contribution of National Laboratory of the Health (LNS) technical staff in Bamako.

## References

[1] World Health Organization (2017) World Malaria Report 2017. World Health Organization, Geneva.

[2] Tchicaya, A.-F., *et al.* (2014) Impacts professionnels et économiques du paludisme à Plasmodium falciparum sur une entreprise du secteur privé en Côte d'Ivoire. *Archives des Maladies Professionnelles et de l'Environnement*, **75**, 406-411. https://doi.org/10.1016/j.admp.2014.02.005

[3] Enquête sur les indicateurs du paludisme au Mali (EIPM) (2015) Institut National de la Recherche en Santé Publique (INRSP) Ministère de la Santé Publique et de l'Hygiène Publique Bamako, Mali. ICF international Rockville, Maryland.

[4] Dondorp, A., Nosten, F., Stepniewska, K., Day, N. and White, N. (2005) Artesunate versus Quinine for Treatment of Severe Falciparum Malaria: A Randomised Trial. *Lancet*, **366**, 717-725. https://doi.org/10.1016/S0140-6736(05)67176-0

[5] Dondorp, A.M., *et al.* (2010) AQUAMAT Group: Artesunate versus Quinine in the Treatment of Severe Falciparum Malaria in African Children (AQUAMAT): An Open-Label, Randomised Trial. *Lancet*, **376**, 1647-1657. https://doi.org/10.1016/S0140-6736(10)61924-1

[6] Cramer, *et al.* (2011) Treatment of Imported Severe Malaria with Artesunate Instead of Quinine—More Evidence Needed? *Malaria Journal*, **10**, 256. https://doi.org/10.1186/1475-2875-10-256

[7] Rolling, *et al.* (2013) Artesunate versus Quinine in Thetreatment of Severe Imported Malaria: Comparative Analysis of Adverseevents Focussing on Delayed Haemolysis. *Malaria Journal*, **12**, 241. https://doi.org/10.1186/1475-2875-12-241

[8] World Health Organization (2016) World Malaria Report 2016. ISBN 978-92-4-151171-1, World Health Organization, Geneva.

[9] World Health Organization (2009) World Health Organization: Global Antimalarial Drug Policies Database—WHO African Region. World Health Organization, Geneva.

[10] Achan, *et al.* (2011) Quinine, an Old Anti-Malarial Drug in a Modern World: Role

in the Treatment of Malaria. *Malaria Journal*, **10**, 144. https://doi.org/10.1186/1475-2875-10-144

[11] Klein, E.Y. (2013) Antimalarial Drug Resistance: A Review of the Biology and Strategies to Delay Emergence and Spread. *International Journal of Antimicrobial Agents*, **41**, 311-317. https://doi.org/10.1016/j.ijantimicag.2012.12.007

[12] Onwujekwe, O., Kaur, H., Dike, N., Shu, E., Uzochukwu, B., Hanson, K., Okoye, V. and Okonkwo, P. (2009) Quality of Anti-Malarial Drugs Provided by Public and Private Healthcare Providers in South-East Nigeria. *Malaria Journal*, **8**, 22. https://doi.org/10.1186/1475-2875-8-22

[13] Lon, C.T., Tsuyuoka, R., Phanouvong, S., Nivanna, N., Socheat, D., Sokhan, C., Blum, N., Christophel, E.M. and Smine, A. (2006) Counterfeit and Substandard Antimalarial Drugs in Cambodia. *Transactions of the Royal Society of Tropical Medicine and Hygiene*, **100**, 1019-1024. https://doi.org/10.1016/j.trstmh.2006.01.003

[14] Brower, V. (2017) Counterfeit and Substandard Malaria Drugs in Africa. *The Lancet Infectious Diseases*, **17**, 1026-1027.

[15] World Health Organization (1997) Quality Assurance of Pharmaceutical: A Compendium of Guidelines and Related Materials: Volume 1.

[16] Dowell, F.E., Maghirang, E.B., Fernandez, F.M., Newton, P.N. and Green, M.D. (2008) Detecting Counterfeit Antimalarial Tablets by Near-Infrared Spectroscopy. *Journal of Pharmaceutical and Biomedical Analysis*, **48**, 1011-1014. https://doi.org/10.1016/j.jpba.2008.06.024

[17] Kouakou, A.K., Soro, A.P., Taky, A.K., Patrice, K. and Zoueu, J.T. (2017) Multi-Spectral and Fluorescence Imaging in Prevention of Overdose of Herbicides: The Case of Maize. *Spectral Analysis Reviews*, **5**, 11-24. https://doi.org/10.4236/sar.2017.52002

[18] (2014) MATLAB and Statistics Toolbox Release. The Math Works, Inc., Natick, MA, USA.

[19] Kishor V.P., Amod, S.P., *et al.* (2015) Application of Area under Curve Technique for UV-Spctrophotometric Determination of Benazepril in Bulk and Tablet Formulation. A*nalytical Chemistry: An Indian Journal*, **15**, 35-38.

[20] Khalid, A.M, Ebrahim, A., *et al.* (2017) Simultaneous Equation and Area under the Curve Spectrophotometric Methods for Estimation of Cefaclor in Presence of Its Acid Induced Degradation Product; a Comparative Study. *Future Journal of Pharmaceutical Sciences*, **3**, 163-167. https://doi.org/10.1016/j.fjps.2017.06.001

[21] Moffa, A.C., Osselton, M.D. and Widdop, B. (2011) Clarke's Analysis of Drugs and Poisons in Pharmaceuticals, Body Fluids and Postmortem Material. 4th Edition, Pharmaceutical Press, London, England.

[22] Irvin, J.L. and Irvin, E.M. (1948) A Fluorometric Method for the Determination of Pamaquine, SN-13276, and SN-3294. *The Journal of Biological Chemistry*, **174**, 589-596.